\documentclass [12pt]{article}

\begin{document}

\begin{center}
\large\textbf{Spectroscopy of the Einstein-Maxwell-Dilaton-Axion black hole}
\end{center}

\begin{center}
Deyou Chen$^{a}$\footnote{ E-mail: \underline
{ruijianchen@gmail.com} } and Haitang Yang$^{b}$\footnote{ E-mail:
\underline {hyanga@uestc.edu.cn} }
\end{center}

\begin{center}
$^{a}$Institute of Theoretical Physics, China West Normal
University, Nanchong, 637002, China

$^{b}${Department of Physics,\\
University of Electronic Science and Technology of China,\\
Chengdu, 610054, China}
\end{center}

\textbf{Abstract:} The entropy spectrum of a spherically symmetric black
hole was derived via the Bohr-Sommerfeld quantization rule in Majhi and
Vagenas's work. Extending this work to charged and rotating black holes,
we quantize the horizon area and the entropy of an
Einstein-Maxwell-Dilaton-Axion (EMDA) black hole via
the Bohr-Sommerfeld quantization rule and the adiabatic invariance. The
result shows the area spectrum and the entropy spectrum are respectively
equally spaced and independent on the parameters of the black hole.

\vspace*{1.0ex}
\textbf{1. Introduction}
\vspace*{1.0ex}

It is widely believed that black holes have discrete horizon area
spectra. The horizon area was first quantized in Bekenstein's work
by adding a particle into the black hole and the discrete spectrum
was derived as \cite{JDB}

\begin{equation}
A_n = 8\pi nl_p^2 , \label{eq:1}
\end{equation}

\noindent
where $l_p $ is the Planck length and $n = 1,2,3 \cdots $. This
implies that the area spectrum is equally spaced and the minimal
spacing is $8\pi l_p^2 $.

One popular method to quantize the horizon area was first put
forward by Hod \cite{SH}. In this method, quasinormal modes (QNMs)
were needed. Using the quasinormal mode frequency and the Bohr's
correspondence principle, he proposed that the area spacing was
$\Delta A = 4l_p^2 \ln 3$ and related to the real part of the modes.
The value of this area spacing is in consistence with both the
Boltzmann-Einstein formula in statistics and the area-entropy
thermodynamic relation for black holes. However, there are some
problems needed to be solved. First, the factor $4\ln 3$
associated with the real part is not a universal value. In
Ref. \cite{MN}, the generic spin-$j$ perturbations were investigated.
It showed that the leading asymptotic value of the quasinormal
mode frequencies is related to spin perturbations and given by
$e^{8\pi M\omega _n } = - \left( {1 + 2\cos \pi j} \right)$, where
$\omega _n = \omega _R + i\omega _I $. When $j = 0$, $j = 1$ and
$j = 2$, it describes the cases of scalar perturbation, vector
perturbation and gravitational perturbation, respectively. It can
be found that the real part would vanish for certain values of $j$.
The research on the asymptotic behavior of the quasinormal mode
frequencies of gravitational perturbations furthermore exhibits this
problem \cite{MN,BK,BCKO}. Moreover, he only considered the case
from the ground state to a state with large $n$. The result would be
changed when one considers two arbitrary states. Black holes
perturbed by exterior fields can be seen as damped oscillators.
The frequency of a damped oscillator is its physical frequency and
is related to both the real part and the imaginary part. Therefore
the physical frequency of the perturbed black hole also has this
property and is

\begin{equation}
\omega = \sqrt {\omega _R^2 + \omega _I^2 } .\label{eq:2}
\end{equation}

\noindent This view was proved in Maggiore's work \cite{MM}.
Applying this new explanation to the quantization of the horizon
area of spherically symmetric black holes, he found that the area
spacing is $8\pi l_p^2 $. This value is different from that
derived by Hod, but in consistence with Bekenstein's result.
Subsequently people have applied this new explanation to quantize
the horizon area of other black holes and all of the results show
the validity of this mode \cite{ECV,KPS,KR,SF,AJM,ALO,WLLR,WLYZ,
LXL,GPS,CYZ,BCS,SV}.

The entropy/area quantization via tunneling mechanism was proposed
in the work of Majhi and his collaborators \cite{BMV1,BRM,BMV2}.
From the Hawking radiation as tunneling effect, the entropy/area spectra
were obtained. The result showed the minimal interval of entropy spectrum
is $1$, which is different from Bekenstein and Maggiore's results.
The area spectra can be obtained from the relation between the entropy
and the area, while the relation is different in different gravity theories.

Recently, the work of Majhi and Vagenas has showed that the black
hole entropy could be quantized without QNMs \cite{MV}. In their
work, the black hole is non-extreme and the horizon is seen as an
adiabatic invariant. From the Hamilton function, they quantized
the entropy of a spherically symmetric black hole via the
Bohr-Sommerfeld quantization rule. The entropy spectrum was derived
and equally spaced, which is full in consistence with that derived by
Maggiore and Bekenstein. One interesting point is that the QNMs
didn't appear in their work. Subsequently, this work has been
extended and people have quantized the entropy via the periodicity
of the particle's wave function \cite{ZLL,AL}.

In this paper, we extend this work to the charged and rotating
black holes. We quantize the horizon area and the entropy of an
Einstein-Maxwell-Dilaton-Axion (EMDA) black hole by the Bohr-Sommerfeld
quantization rule and the adiabatic invariance. The result shows both
 the area spectrum and the entropy spectrum are equally spaced.

The rest is organized as follows. In Sect. 2, the EMDA spacetime
is reviewed. In Sect. 3, we first derive the Hamiltonian of the
EMDA black hole system, and then quantize the horizon area and the
entropy via the Bohr-Sommerfeld quantization rule. In Sect. 4 we
offer a summary and discussion.

\vspace*{2.0ex}
\textbf{2. The EMDA spacetime}
\vspace*{1.0ex}

The metric of the EMDA black hole is given by \cite{GGK}

\begin{eqnarray}
ds^2 & = &- \frac{\Delta - a^2\sin ^2\theta }{\sum }dt^2 + \frac{\sum }{\Delta
}dr^2 + \sum d\theta ^2 + \frac{2a\sin ^2\theta \left( {r^2 + 2br + a^2 -
\Delta } \right)}{\sum }dtd\phi\nonumber\\
 &&+ \frac{\left( {r^2 + 2br + a^2} \right)^2 - \Delta a^2\sin ^2\theta }{\sum
}\sin ^2\theta d\phi ^2, \label{eq:3}
\end{eqnarray}

\noindent with the electromagnetic vector potential

\[
{A}'_\mu = {A}'_t dt + {A}'_\phi d\phi = \frac{Qr}{\sum }dt - \frac{Qra\sin
^2\theta }{\sum }d\phi ,
\]

\noindent
where

\begin{eqnarray*}
r_\pm & = & M\pm \sqrt {M^2 - a^2} ,\\
\sum & = & r^2 + 2br + a^2\cos ^2\theta ,\\
\Delta & = & r^2 - 2Mr + a^2 = \left( {r - r_ + } \right)\left( {r
- r_ - } \right).
\end{eqnarray*}

\noindent $r_ + \left( {r_ - } \right)$ is the out (inner)
horizon, $M$ and $a$ are the physical mass and the angular
momentum per unit mass respectively, and $b$ is defined as $b =
\raise0.7ex\hbox{${Q^2}$} \!\mathord{\left/ {\vphantom {{Q^2}
{2M}}}\right.\kern-\nulldelimiterspace}\!\lower0.7ex\hbox{${2M}$}$.
The ADM mass and the angular momentum are given by $M_A = M + b$,
$J = \left( {M + b} \right)a$.  The thermodynamic properties of
the EMDA black hole have been deeply discussed. The horizon area,
entropy, angular velocity and Hawking temperature are respectively

\begin{eqnarray}
A & = & 4\pi \left( {r_ + ^2 + 2br_ + + a^2} \right),
\nonumber\\
S & = & \pi \left( {r_ + ^2 + 2br_ + + a^2} \right),
\nonumber \\
T & = & \frac{r_ + ^2 - a^2}{4\pi r_ + \left( {r_ + ^2 + 2br_ + + a^2}
\right)},
\nonumber\\
\Omega _ + & = & \frac{a}{r_ + ^2 + 2br_ + + a^2}. \label{eq:4}
\end{eqnarray}

\noindent In the extreme case, the out horizon and the inner
horizon coincide with each other. The surface gravity vanishes in
this situation and the black hole entropy is $S = 2\pi \left( {M +
b} \right)r_ + = 2\pi J$.

\vspace*{2.0ex}
\textbf{3. Spectroscopy of the EMDA black hole}
\vspace*{1.0ex}

In this paper, we quantize the horizon area of the EMDA black hole
by combining the Bohr-Sommerfeld quantization rule and the
adiabatic invariance. There is a frame-dragging effect of the
coordinate system in the EMDA spacetime and the matter field in
the ergosphere near the horizon must be dragged. It is convenient
to investigate the black hole's properties in the dragging
coordinate system. Thus we perform the dragging coordinate
transformation \cite{CY}

\begin{eqnarray}
d\phi =  -\frac{g_{03}}{g_{33}}dt
= \frac{a\left( {r^2 + 2br + a^2 - \Delta }
\right)}{\left( {r^2 + 2br + a^2} \right)^2 - \Delta a^2\sin
^2\theta }dt. \label{eq:5}
\end{eqnarray}

\noindent
Inserting Eq. (\ref{eq:5}) into the metric (\ref{eq:3}) yields

\begin{eqnarray}
ds^2 = - \frac{\Delta \sum }{\left( {r^2 + 2br + a^2} \right)^2 -
\Delta a^2\sin ^2\theta }dt^2 + \frac{\sum }{\Delta }dr^2 + \sum d\theta ^2.
\label{eq:6}
\end{eqnarray}

\noindent Now the electromagnetic vector potential in three
dimensional spacetime is expressed as ${A}_\mu = \left( {{A}_t
,0,0} \right)$, with ${A}_t =  \left( {r^2 + 2br + a^2} \right)Qr
[\left( {r^2 + 2br + a^2}\right)^2 - \Delta a^2\sin ^2\theta
]^{-1}$. The Bohr-Sommerfeld quantization rule tells us

\begin{equation}
\int {p_i dq_i } = nh,\label{eq:7}
\end{equation}

\noindent
where $n = 1,2,3 \cdots $. To quantize the horizon area,
we first Euclideanize the EMDA metric. In Ref. \cite{ZZM}, the
metric of a rotating spacetime is Euclideanized via the
transformation $t \to - i\tau $ and $a \to ia$. In the dragging
coordinate system, the Euclideanized EMDA metric is obtained by a
transformation $t \to - i\tau $ in the metric (\ref{eq:6}) and
takes on the form as

\begin{eqnarray}
ds^2 =  \frac{\Delta \sum }{\left( {r^2 + 2br + a^2} \right)^2 -
\Delta a^2\sin ^2\theta }d\tau^2 + \frac{\sum }{\Delta }dr^2 + \sum d\theta ^2.
\label{eq:8}
\end{eqnarray}

\noindent
Now the corresponding electromagnetic vector potential is  $A_\mu =
\left( {A_\tau ,0,0} \right)$.

The horizon can be seen as an adiabatic invariant when a black
hole is non-extreme. Therefore, we can get

\begin{equation}
\int {p_i dq_i } = \int {\int_0^H {\frac{d{H}'}{\dot {q}_i }dq_i } } =
\int\limits {\int_0^H {d{H}'d\tau } } + \int {\int_0^H
{\frac{d{H}'}{\dot {x}_a }dx_a } } ,
\label{eq:9}
\end{equation}

\noindent where $\tau $ is the Euclidean time, which has a
periodicity $\raise0.7ex\hbox{${2\pi }$} \!\mathord{\left/
{\vphantom {{2\pi } \kappa
}}\right.\kern-\nulldelimiterspace}\!\lower0.7ex\hbox{$\kappa $}$
and $\kappa $ is the surface gravity, and $ x_a$'s denote the
generalized space coordinates and $a=1,2,3$. $H$ is the Hamilton function
of the black hole's system and satisfies $H = \int\limits_{\tau _i
}^{\tau _f } {Ld\tau } $, with $L$ being the Lagrangian. When
investigate a particle tunnelling through the horizon, one has to
take into account the effect of the electromagnetic field.
Therefore, the gravitational system consists of the black hole and
the outside electromagnetic field. The Lagrangian is composed of
the part from the matter field and that from the electromagnetic
field. $L_e = - 1 / 4F_{\mu \nu } F^{\mu \nu }$ is the Lagrangian
function of the generalized coordinates $A_\mu = \left( {A_\tau
,0,0} \right)$. From the function, we find that $A_\tau $ and
$\phi $ are cyclic coordinates. To eliminate the degrees of
freedom corresponding to $\phi $ and $A_\tau $, the Hamilton
function should be written as

\begin{equation}
H = \int\limits_{\tau _i }^{\tau _f } {\left( {L - P_\phi \dot {\phi } - P_A
\dot {A}_\tau } \right)d\tau } = \int\limits_{r_i }^{r_f }
{\int\limits_0^{P_r } {d{P}'_r dr} } - \int\limits_{\phi _i }^{\phi _f }
{\int\limits_0^{P_\phi } {d{P}'_\phi d\phi } } - \int\limits_{A_{ti}
}^{A_{tf} } {\int\limits_0^{P_A } {d{P}'_A dA_\tau } } .
\label{eq:10}
\end{equation}

\noindent  $P_r $, $P_\phi $ and $P_A $ are canonical momenta of
$r$, $\phi $ and $A_\tau $. $r_i $ and $r_f $ are locations of the
outer horizons before and after the emission of a particle,
respectively. To further proceed, we introduce the Hamilton
canonical equations. They are expressed as

\begin{eqnarray}
\dot {r} & = & \left. {\frac{dH}{dP_r }} \right|_{\left( {r;\phi ,P_\phi ;A_\tau
,P_A } \right)} ,
\quad
\left. {dH} \right|_{\left( {r;\phi ,P_\phi ;A_\tau ,P_A } \right)} =
dM' ;\nonumber \\
\dot {\phi } & = & \left. {\frac{dH}{dP_\phi }} \right|_{\left( {\phi ;r,P_r
;A_\tau ,P_A } \right)} ,
\quad
\left. {dH} \right|_{\left( {\phi ;r,P_r ;A_\tau ,P_A } \right)} = {\Omega
}'_\tau dJ';
\nonumber \\
\dot {A}_\tau&  = & \left. {\frac{dH}{dP_A }} \right|_{\left( {A_\tau ;\phi
,P_\phi ;r,P_r } \right)} ,
\quad
\left. {dH} \right|_{\left( {A_\tau ;\phi ,P_\phi ;r,P_r ;} \right)} =
{A}'_\tau dQ'; \label{eq:11}
\end{eqnarray}

\noindent in which ${A}'_\tau $ and ${\Omega }'_\tau $ are the
electromagnetic potential and the angular velocity with the
emission of particles. When the emitted particle is massless, the
outgoing path is the radial null geodesic $\dot {r} =
\raise0.7ex\hbox{${dr}$} \!\mathord{\left/ {\vphantom {{dr} {d\tau
}}}\right.\kern-\nulldelimiterspace}\!\lower0.7ex\hbox{${d\tau
}$}$ \cite{PW}. The outgoing path is the phase velocity ($\dot {r}
= v_p )$ of the particle when the particle is charged and massive
\cite{JW,ZZ}. Inserting Eq. (\ref{eq:11}) into Eq. (\ref{eq:10})
yields

\begin{eqnarray}
H & = & \int\limits_{\tau _i }^{\tau _f } {\left( {L - P_\phi \dot {\phi } - P_A
\dot {A}_\tau } \right)d\tau } \nonumber \\
& = &\int\limits_{\tau _i }^{\tau _f } {\left( {\int\limits_0^H {\left. {d{H}'}
\right|_{\left( {r;\phi ,P_\phi ;A_\tau ,P_A } \right)} - \int\limits_0^H
{\left. {d{H}'} \right|_{\left( {r;\phi ,P_\phi ;A_\tau ,P_A } \right)} -
\int\limits_0^H {\left. {d{H}'} \right|_{\left( {A_\tau ;\phi ,P_\phi ;r,P_r
;} \right)} } } } } \right)d\tau } \nonumber \\
& = & \int\limits_{\tau _i }^{\tau _f } {\int\limits_0^H {d{H}'d\tau } }.
\label{eq:12}
\end{eqnarray}

\noindent
Combining Eqs.(\ref{eq:9}) and (\ref{eq:12}), we can rewrite the adiabatic
invariant as

\begin{eqnarray}
\int {p_i dq_i } & = & 2\int{\int\limits_0^H {d{H}'d\tau } } \nonumber\\
& = & 2\int {\left( {\int\limits_0^H {\left. {d{H}'}
\right|_{\left( {r;\phi ,P_\phi ;A_\tau ,P_A } \right)} - \int\limits_0^H
{\left. {d{H}'} \right|_{\left( {r;\phi ,P_\phi ;A_\tau ,P_A } \right)} -
\int\limits_0^H {\left. {d{H}'} \right|_{\left( {A_\tau ;\phi ,P_\phi ;r,P_r
;} \right)} } } } } \right)d\tau } \nonumber \\
& = & 2\int{\left( {\int\limits_0^M {dM' - \int\limits_0^J {{\Omega }'_ {\tau} dJ'}
- \int\limits_0^Q {{A}'_ {\tau} dQ'} } } \right)d\tau } .
\label{eq:13}
\end{eqnarray}

\noindent
Here we only consider the outgoing path, which implies the half value of
periodicity of the Euclidean time is selected, namely $0 \le \tau \le \frac{\pi}{\kappa}$.
From the first law of thermodynamics of the EMDA black hole

\begin{eqnarray}
dM = TdS + \Omega  dJ + \Phi dQ, \label{eq:14}
\end{eqnarray}

\noindent
where $\Omega $ and $\Phi$ are the angular velocity and the electromagnetic potential
at the horizon, we finish the integral and get

\begin{eqnarray}
\int {p_i dq_i }  =  2\int\limits_0^{\frac{\pi}{\kappa}}{\int\limits_0^H {d{H}'d\tau } }
 = 2\int_0^S {\frac{\pi}{\kappa} \cdot Td{S}'} = \hbar S, \label{eq:15}
 \end{eqnarray}

\noindent where the last equality is obtained by a relation
between the surface gravity and the Hawking temperature $T =
\frac{\hbar \kappa}{2\pi} $. Introducing the Bohr-Sommerfeld
quantization rule in Eq. (\ref{eq:7}), we derive the entropy
spectrum as

\begin{equation}
S = 2\pi n. \label{eq:16}
\end{equation}

\noindent
Therefore the minimal interval of the entropy spectrum is $\Delta S =
S_n - S_{n - 1} = 2\pi $. It shows the entropy spectrum is equally spaced
and independent on the parameters of the EMDA black hole. From the area-entropy
law $S = \raise0.7ex\hbox{$A$} \!\mathord{\left/ {\vphantom {A {4l_p^2
}}}\right.\kern-\nulldelimiterspace}\!\lower0.7ex\hbox{${4l_p^2 }$}$, the
horizon area spectrum is obtained as

\begin{equation}
A = 8\pi nl_p^2 . \label{eq:17}
\end{equation}

\noindent
This implies the minimal interval of the area spectrum is $\Delta A = 8\pi l_p^2 $.
This value is in consistence with that derived by Bekenstein and Maggiore \cite{JDB,MM}.
So the area spectrum and the entropy spectrum are respectively equally
spaced and independent on the parameters of the EMDA black hole.

\vspace*{2.0ex}
\textbf{4. Discussion and Conclusion}
\vspace*{1.0ex}

In this paper, extending the work of Majhi and Vagenas to the charged and
rotating black holes, we  quantized the horizon area and the entropy of
the EMDA black hole via the Bohr-Sommerfeld quantization rule and the
adiabatic invariance. The area spectrum and the entropy spectrum
were derived and are respectively equally spaced, which are independent on
the parameters of the EMDA black hole. This result is in consistence
with that obtained by Maggiore and that derived by Bekenstein. In the
investigation, the area spectrum was derived by the area-entropy relation
$S = \frac{A}{4l^2_p}$. However, if the relation does not satisfy
$S = \frac{A}{4l^2_p}$, the area spectrum would be changed. This was
addressed in Majhi and Vagenas's work \cite{MV}.

\vspace*{3.0ex}
{\bf Acknowledgements}
\vspace*{1.0ex}

This work is supported in part by the Natural Science Foundation
of China (Grant No.11178018 and No. 11175039).


\bigskip

\begin{thebibliography}{99}

\small

\bibitem{JDB}
J.~D.~Bekenstein, Lett. Nuovo Cimento {\bf11}, 467 (1974).

\bibitem{SH}
S.~Hod, Phys. Rev. Lett., {\bf81}, 4293 (1998).

\bibitem{MN}
L. Motl, A. Neitzke, Adv. Theor. Math. Phys., {\bf7} 307 (2003)

\bibitem{BK}
E. Berti, K. D. Kokkotas, Phys. Rev. D, {\bf68} 044027 (2003).

\bibitem{BCKO}
E. Berti, V. Cardoso, K. D. Kokkotas, H. Onozawa, Phys. Rev. D, {\bf68} 124018 (2003).

\bibitem{MM}
M.~Maggiore, Phys. Rev. Lett., {\bf100}, 141301 (2008).

\bibitem{ECV}
E.~C.~Vagenas, J. High Energy Phys. {\bf0811}, 073 (2008).

\bibitem{KPS}
D.~Kothawala, T.~Padmanabhan, S.~Sarkar, Phys. Rev. D, {\bf78},
104018 (2008).

\bibitem{KR}
K.~Ropotenko, Phys. Rev. D, {\bf80}, 044022 (2009).

\bibitem{SF}
S.~Fernando, Phys. Rev. D, {\bf79}, 124026 (2009).

\bibitem{AJM}
A.~J.~M.~Medved, Class. Quant. Grav. {\bf25}, 205014 (2008).
M.~R.~Setare, E.~C.~Vagenas, Mod. Phys. Lett. A, {\bf20}, 1923 (2005).

\bibitem{ALO}
A.~Lopez-Ortega, Phys. Lett. B, {\bf682}, 85 (2009).

\bibitem{WLLR}
S.~W.~Wei, R.~Li, Y.~X.~Liu, J.~R.~ Ren, J. High Energy Phys., {\bf0903},
076 (2009).

\bibitem{WLYZ}
S.~W.~Wei, Y.~X.~Liu, K.~Yang, Y.~Zhong, Phys. Rev. D, {\bf81}, 104042 (2010).

\bibitem{LXL}
W.~Li, L.~Xu, J.~Lu, Phys. Lett. B {\bf676}, 177 (2009).

\bibitem{GPS}
P. Gonzalez, E. Papantonopoulos, J. Saavedra, J. High Energy Phys., 08 050 (2010).

\bibitem{CYZ}
D.~Chen, H.~Yang, X.~T.~Zu, Eur. Phys. J. C, {\bf69}, 289 (2010).

\bibitem{BCS}
E.~Berti, V.~Cardoso, A.~O.~Starinets, Quasinormal modes of black
holes and black branes. arXiv:0905.2975 [gr-qc]

\bibitem{SV}
M.~R.~Setare, E.~C.~Vagenas, Mod. Phys. Lett. A 20 1923 (2005).

\bibitem{BMV1}
R.~Banerjee, B.~R.~Majhi, E.~C.~Vagenas, Phys. Lett. B, {\bf686}, 279
(2010).

\bibitem{BRM}
B.~R.~Majhi, Phys. Lett. B, {\bf686} 49 (2010).

\bibitem{BMV2}
R.~Banerjee, B.~R.~Majhi, E.~C.~Vagenas, Euro. phys. Lett., {\bf92}
20001 (2010).

\bibitem{MV}
B.~R.~Majhi, E.~C.~Vagenas, Phys. Lett. B, {\bf701}, 623 (2011).

\bibitem{ZLL}
X.~X.~Zeng, X.~M.~Liu, W.~B.~Liu, Periodicity and area spectrum of
black holes. arXiv:1203.5947[gr-qc];
X.~X.~Zeng, W.~B.~Liu, Spectroscopy of a Reissner-Nordstrom black
hole via an action variable. arXiv:1204.1699[gr-qc];
X.~M.~Liu, X.~X.~Zeng, W.~B.~Liu, Spectroscopy of the rotating BTZ
black hole via adiabatic invariance. arXiv:1204.1786 [gr-qc].


\bibitem{AL}
A.~Larranaga, Area spectrum of a rotating charged black hole solution
of heterotic string theory. arXiv:1204.0851 [gr-qc].

\bibitem{GGK}
A.~A.~Garcia, D.~V.~Gal'tsov, O.~V.~Kechkin, Phys. Rev. Lett., {\bf74},
1276 (1995).

\bibitem{CY}
D.~Chen, S.~Z.~Yang, Int. J. Mod. Phys. D, {\bf16},1285 (2007).

\bibitem{ZZM}
Z. Z. Ma, Phys. Rev. D, {\bf67} 024027 (2003).

\bibitem{PW}
M.~K.~Parikh, F.~Wiltzek, Phys. Rev. Lett., {\bf85}, 5042 (2000).

\bibitem{JW}
Q. Q. Jiang, S. Q. Wu, Phys. Lett. B, {\bf635} 51 (2006).

\bibitem{ZZ}
J.~Y.~Zhang, Z.~Zhao, Phys. Lett. B, {\bf638}, 110 (2006).












\end{thebibliography}
\end{document}